\documentclass[11pt]{article}
\usepackage{geometry}                
\usepackage{graphicx}
\usepackage{amsmath}
\usepackage{latexsym}
\usepackage{amssymb}
\newcommand{\bfr}{{\bf r}}

\newcommand{\bfE}{{\bf E}}
\newcommand{\bfB}{{\bf B}}
\newcommand{\bfb}{{\bf b}}

\newcommand{\bfD}{{\bf D}}

\newcommand{\bfA}{{\bf A}}
\newcommand{\bfa}{{\bf a}}
\newcommand{\bfp}{{\bf p}}
\newcommand{\bdr}{{\bf \dot{r}}}
\newcommand{\bdp}{{\bf \dot{p}}}
\newcommand{\fmn}{{F_{\mu \nu}}}
\newcommand{\fmna}{{F^{\ a}_{\mu \nu}}}
\newcommand{\fumn}{{F^{\mu \nu}}}
\newcommand{\lym}{{\mathcal{L}_{\scriptscriptstyle Y\! M}}}
\newcommand{\iym}{{I_{\scriptscriptstyle Y\! M}}}
\newcommand{\fdast}{{^*F}_{\mu \nu}}
\newcommand{\fdasta}{{^*F}^{\ a}_{\mu \nu}}
\newcommand{\fuast}{{^*F}^{\mu \nu}}
\newcommand{\fuasta}{{^*F}^{\mu \nu \ a}}
\newcommand{\astf}{{^\ast F}}
\newcommand{\varep}{{\varepsilon^{\mu \alpha\beta\gamma}}}
\newcommand{\UU}{{U\, U^{-1}\,}}
\newcommand{\U}{{U^{-1}\,}}

\newcommand{\hatp}{{\hat{\varphi}}}
\newcommand{\doe}{DE-FC02-94ER40818.}

\newcommand{\jmu}{{\langle j^\mu_5\rangle}}
\newcommand{\Rxy}{{R(x, y; \mu)}}

\let\vec\boldsymbol
\numberwithin{equation}{section}

\title{Topological Aspects of Gauge Theories \footnote{This work is supported in part by funds provided by the U.S. Department of Energy (D.O.E) under cooperative research agreement \doe}} 
\author{R. Jackiw\\
\it \small Center for Theoretical Physics\\
\it \small Massachusetts Institute of Technology\\
\it \small Cambridge, MA 02139-4307}
\date{\small MIT-CTP-3593}                                           

\begin{document}

\maketitle
\begin{abstract}
To appear in Encyclopedia of Mathematical Physics, published by  Elsevier in early 2006. Comments/corrections welcome. The article surveys topological aspects in gauge theories.
\end{abstract}
\section{Introduction}
Classical fields that enter a classical field theory provide a mapping from the ``base" manifold on which they are defined (space or space-time) to a ``target" space over which they range. The base and target spaces, as well as the map, may possess non-trivial topological features, which affect the fixed-time description and the temporal evolution of the fields, thereby influencing the physical reality that these fields describe. Quantum fields of a quantum field theory are operator-valued distributions whose relevant topological properties are obscure. Nevertheless, topological features of the corresponding classical fields are important in the quantum theory for a variety of reasons: (i) Quantized fields can undergo local (space-time dependent) transformations (gauge transformations, coordinate diffeomorphisms) that involve classical functions whose topological properties determine the allowed  quantum field theoretic structures. (ii) One formulation of the quantum field theory uses a functional integral over classical fields, and  classical topological features become relevant. (iii) Semi-classical (WKB) approximations to the quantum theory rely on classical dynamics, and again classical topology plays a role in the analysis.

Topological effects of gauge fields in quantum theory were first appreciated by Dirac in his study of the quantum mechanics for (hypothetical) magnetic point-monopoles. Although here one is not dealing with a field theory, the consequences of his analysis contain many features that were later encountered in field theory models. 

The Lorentz equations of motion for a charged (e) massive $(M)$ particle in a monopole magnetic field $(\bfB = m \bfr/r^3)$ are unexeptional, 
\begin{subequations}\label{eqone1}
\begin{eqnarray}
\bdr = \frac{\bf p}{M} \qquad \quad &\label{eqone1a}\\
\bdp= \frac{e}{M} \, \bfp \times \bfB  &\quad (c = 1)\label{eqone1b},
\end{eqnarray}
\end{subequations}
and completely determine classical dynamics. But knowledge of the Lagrangian $L$ and of the action $I$ -- the time integral of  $L:I = \int d t L$ -- is further needed for quantum mechanics, either in its functional integral formulation or in its Hamiltonian formulation, which requires the canonical momentum $\vec\pi \equiv \partial L/\partial \bdr$. 
The Lorentz-force action is expressed in terms of the vector potential $\bfA, \bfB = \vec \nabla \times
 {\bf A}\negthickspace: I_{\text{\tiny Lorentz}} = e \int d t \bdr \cdot \bfA = e \int d \bfr \cdot \bfA$. The magnetic monopole vector potential is necessarily singular because $\vec\nabla \cdot \bfB = 4\pi m \delta^3 (\bfr) \ne 0$. 
The singularity (Dirac string) can be moved, but not removed, by gauge transformations, which also are singular, and do not leave the Lorentz action invariant. 
Noninvariance of the action can be tolerated provided its change is an integral multiple of  $2\pi$, since the functional integrand  involves  $exp\, i\,  I, (\hbar = 1)$. The quantal requirement, which is not seen in the equations of motion, is met when 
\begin{equation}
eg = N/2.
\label{eqone2}
\end{equation}
The topological background to this (Dirac) quantization condition is the fact that $\Pi_1 \big(U(1)\big)$ is the group of integers, {\it i.e.} the map of the unit circle into the gauge group, here  $U(1)$, is classified by integers.

Further analysis shows that only point magnetic sources can be incorporated in particle quantum mechanics, which is governed by the particle Hamiltonian $H=\frac{\bfp^2}{2M}$ (magnetic fields do no work and are not seen in $H$). Quantum Lorentz equations are regained by commutation with $H$: $\dot{\bfr} = i [H, \bfr], \dot{\bfp} = i [H, \bfp]$, provided
\begin{subequations}\label{eqone3}
\begin{alignat}{2}
&i\, [r^i, r^j] = 0,  \label{eqone3a}\\
&i\, [p^i, r^j] = \delta^{ij}, \label{eqone3b}\\
&i\, [p^i, p^j] = - e \varepsilon^{ijk}\, B^k. \label{eqone3c}
\end{alignat}
\end{subequations}
But (\ref{eqone3c}) implies that the Jacobi identity is obstructed by magnetic sources $\vec\nabla \cdot \bfB \ne 0$.
\begin{equation}
\varepsilon^{ijk}\, [p^i, p^j p^k] = e \vec\nabla \cdot \bfB
\label{eqone4}
\end{equation}
This obstruction is better understood by examining the unitary operator $U(\bfa) \equiv exp\, (i\, \bfa \cdot \bfp)$, which according to (\ref{eqone3b}) implements finite translations of  $\bfr\,  \text{by}\, \bfa$. The commutator algebra (\ref{eqone3}) and the failure of the Jacobi identity (\ref{eqone4}) imply that these operators do not associate. 
Rather one finds
\begin{equation}
U(\bfa_1) \, \Big(U(\bfa_2) U(\bfa_3)\Big) = e^{i\Phi} \Big(U(\bfa_1)\, U(\bfa_2)\Big)\, U(\bfa_3),
\label{eqone5}
\end{equation}
where $\Phi = e \int d^3 x \, \vec\nabla \cdot \bfB$ is the total flux emerging from the tetrahedron formed from the three vectors $\bfa_i$ with vertex at $\bfr$. see Figure 1.
But quantum mechanics realized by linear operators acting on a Hilbert space requires that operator multiplication be associative. This can be achieved, in spite of (\ref{eqone5}), provided $\Phi$ is an integral multiple of $2\pi$, hence invisible in the exponent. This then needs that (i) $\vec\nabla \cdot \bfB$ be localized at points, so that the volume integral of  $\vec\nabla \cdot \bfB$ retain integrality for arbitrary $\bfa_i$ and (ii) the strengths of the localized poles obey Dirac quantization.
The points at which $\vec\nabla \cdot \bfB$ is localized can now be removed from the manifold and the Jacobi identity is regained.  The above argument, which re-derives Dirac's quantization, makes no reference to gauge variance of magnetic potentials.\\
\begin{figure}[t]
   \centering
   \includegraphics{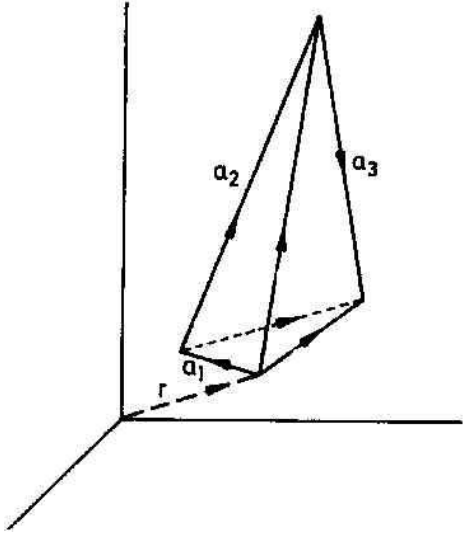} 
   \caption{Tetrahedron pierced by magnetic flux that obstructs associativity.}
    \label{tetra}
\end{figure}
In the remainder we shall discuss related phenomena for selected gauge field theories in 4, 3 and 2 dimensions that describe actual physical events ocurring in Nature. We shall encounter in generalized form, analogs to the above quantum mechanical system.

Some definitions and notational conventions: Non-Abelian gauge potentials  $A^a_\mu$ carry a space-time index $(\mu)$ [metric tensor $g_{\mu \nu} = \text{diag} (1, - 1, \ldots)]$ and an adjoint group index (a). When contracted with anti-Hermitian matrices $T_a$ that represent the group's Lie algebra (structure constraints $f^{\ \, c}_{a b}$)
\begin{equation}
[T_a, T_b] = f^{\ \, c}_{a b}\, T_c,
\label{eqone6}
\end{equation}
they become Lie-algebra valued.
\begin{equation}
A_\mu \equiv A^a_\mu \, T_a
\label{eqone7}
\end{equation}
Gauge transformations transform $A_\mu$ by group elements $U$.
\begin{subequations}\label{eqone8}
\begin{equation}
A_\mu \to A^U_\mu \equiv U^{-1}\, A_\mu \, U + U^{-1} \, \partial_\mu \, U
\label{eqone8a}
\end{equation}
For infinitesimal gauge transformations, $U\approx I + \lambda,  \lambda \equiv \lambda^a\, T_a$; this leads to the covariant derivative $D_\mu$.
\begin{alignat}{2}
& A_\mu \to \ A_\mu + \partial_\mu \, \lambda + [A_\mu, \lambda] \equiv A_\mu + D_\mu \lambda \nonumber \\
& A^a_\mu \to \ A^a_\mu + \partial_\mu \, \lambda^a + f^{\ \, a}_{b c} \, A^b_\mu \, \lambda^c \equiv A^a_\mu + (D_\mu \lambda)^a
\label{eqone8b}
\end{alignat}
\end{subequations}
(In a quantum field theory $A_\mu$ becomes an operator but the gauge transformations $U, \lambda$ remain c-number functions.) The field strength $\fmn$
\begin{subequations}\label{eqone9}
\begin{equation}
\fmn = \partial_\mu \, A_\nu - \partial_\nu \, A_\mu + [A_\mu, A_\nu],
\label{eqone9a}
\end{equation}
also is given by
\begin{equation}
[D_\mu, D_\nu]... = [\fmn, ...],
\label{eqone9b}
\end{equation}
\end{subequations}
(coupling strength $g$ has been scaled to unity). The definition (\ref{eqone9}) implies the Bianchi identity. 
\begin{equation}
D_\mu F_{\nu \omega} + D_\omega \fmn + D_\nu \, F_{\omega \mu}= 0
\label{eqone10}
\end{equation}
$\fmn$ is gauge covariant
\begin{subequations}\label{eqone11}
\begin{equation}
\fmn \to \fmn^U = U^{-1} \, \fmn \, U,
\label{eqone11a}
\end{equation}
or infinitesimally
\begin{equation}
\fmn \to \fmn + [\fmn, \lambda].
\label{eqone11b}
\end{equation}
\end{subequations}
In the gauge invariant Yang-Mills action $I_{YM}$, the Yang-Mills Lagrange density $\lym$ is integrated over the base space.
\begin{eqnarray}
\lym &=& \frac{1}{2}\, t r F^{\mu \nu}\, \fmn\nonumber\\
I_{YM} = \int \, \lym &=& \frac{1}{2} \int t r F^{\mu \nu} \fmn
\label{eqone12}
\end{eqnarray}
The trace is evaluated with the convention
\begin{equation}
tr\, T_a\, T_b = -\frac{1}{2} \, \delta_{ab},
\label{eqone13}
\end{equation}
and henceforth there is no distinction between upper and lower group indices.
The Euler-Lagrange condition for  stationarizing $I_{\scriptscriptstyle YM}$ gives the Yang-Mills equation.
\begin{subequations}\label{eqone14}
\begin{equation}
D_\mu \, F^{\mu \nu} = 0
\label{eqone14a}
\end{equation}
Should sources $J^\mu$ be present, (\ref{eqone14a}) becomes
\begin{equation}
D_\mu \, F^{\mu \nu} = J^\nu,
\label{eqone14b}
\end{equation}
\end{subequations}
and $J^\mu$ must be covariantly conserved.
\begin{equation}
D_\nu \, J^\nu = D_\nu \, D_\mu F^{\mu \nu} = -\frac{1}{2} [D_\mu, D_\nu] F^{\mu \nu} = -\frac{1}{2} [F_{\mu \nu}, F^{\mu \nu}]=0
\label{eqone15}
\end{equation}
All this is a non-Abelian generalization of familiar Maxwell electrodynamics.

\section{Gauge Theories in Four Dimensions}\label{Sec:2}

Gauge theories in 4-dimensional space-time are at the heart of the standard particle physics model. Their topological features have physical consequences and merit careful study.

\vskip 8pt
\noindent
\subsection*{A. Yang-Mills Theory}

In four dimensions, we define non-Abelian electric $\bfE^a \text{and magnetic}\, \bfB^a\,  \text{fields}$.
\begin{equation}
E^{ia} = F^a_{0i}, \qquad B^{ia} = - \frac{1}{2} \, \varepsilon^{ijk} \, F^a_{jk}
\label{eqtwo1}
\end{equation}
Canonical analysis and quantization is carried out in the Weyl gauge $(A^a_0 = 0)$, where the Lagrangian and  Hamiltonian (energy) densities read
\begin{eqnarray}
\lym &=& \frac{1}{2} \, (\bfE^a \cdot \bfE^a - \bfB^a \cdot \bfB^a), \label{eqtwo2}\\
\mathcal{H}_{\scriptscriptstyle YM} &=& \frac{1}{2} \, (\bfE^a \cdot \bfE^a + \bfB^a \cdot \bfB^a).
\label{eqtwo3}
\end{eqnarray}
The first term is kinetic, with $\bfE^a = - \partial_t \, \bfA^a$ also functioning as the (negative) canonical momentum $\vec\pi^a$, conjugate to the canonical variable $\bfA^a$; the second magnetic term gives the  potential.  In the Weyl gauge, the theory remains invariant against time-independent gauge transformations.
The time component of equation (\ref{eqone14}) (Gauss law) is absent (because there is no $A^a_0$ to vary); rather it is imposed as a fixed-time constraint on the canonical variables $\bfE^a \text{and}\, \bfA^a$. This regains the Gauss law.
\begin{subequations}\label{eqtwo4}
\begin{equation}
(\vec{\mathcal{D}} \cdot \bfE)^a = 0 \qquad \text{(in the absence of sources)}
\label{eqtwo4a}
\end{equation}

In the quantum theory $\boldsymbol{\mathcal{D}} \cdot \bfE$ annihilates ``physical" states. Explicitly, in a functional Schr\"{o}dinger representation, where states are functionals of the canonical fixed-time variable $\bfA  |\Psi >\to \Psi(\bfA)$, (\ref{eqtwo4a}) requires
\begin{equation}
\Big(\vec{\mathcal{D}} \cdot \frac{\delta}{\delta \bfA}\Big)^a\, \Psi(\bfA) = 0,
\label{eqtwob}
\end{equation}
\end{subequations}
{\it i.e.} physical states must be invariant against infinitesimal gauge transformation, or equivalently, against gauge transformations that are homotopic (continuously deformable) to the identity (so-called ``small" gauge transformations).
\begin{equation}
\Psi (\bfA + \vec{\mathcal{D}}\, \lambda) = \Psi (\bfA)
\label{eqtwo5}
\end{equation}
But homotopically non-trivial gauge transformation functions that cannot be deformed to the identity  (so-called ``large" gauge transformations) may be present. Their effect is not controlled by Gauss' law, and must be discussed separately. 

Fixed-time gauge transformation functions depend on the spatial variable $\bfr: U(\bfr)$.
For a topological classification, we require that $U$ tend to a constant at large $r$. Equivalently we compactify the base space $R^3\, \text{to}\, S^3$. Thus the gauge functions provide a mapping from $S^3$ into the relevant gauge group $G$, and for non-Abelian compact gauge groups such mappings fall into disjoint homotopy classes labeled by an integer winding number $n$: $\Pi^3 (G) = Z$. Gauge functions $U_n$ belonging to different classes cannot be deformed into each other; only those in the ``zero" class are deformable to the identity. An analytic expression for the winding number $\omega\, (U)$ is 
\begin{equation}
\omega\, (U) = \frac{1}{24\pi^2}\, \int d^3 x \, \varepsilon^{ijk}\, t r (U^{-1}\, \partial_i\, U U^{-1}\, \partial_j \, \UU \partial_k\, U).
\label{eqtwo6}
\end{equation}

This is a most important topological entity for gauge theories in 4-dimensional space-time, {\it i.e.}\,in 3-space, and we shall meet it again in a description of gauge theories in 3-dimensional space-time, {\it i.e.} on a plane.\,Various features of $\omega$ expose its topological character: (i) $\omega\, (U)$ does not involve a metric tensor, yet it  is diffeomorphism invariant. (ii) $\omega\, (U)$ does not change under local variations of $U$.
\begin{eqnarray}
\delta\, \omega\, (U) &=& \frac{1}{8\pi^2} \int d^3 x \, \partial_i\, \varepsilon^{ijk} t r (U^{-1} \, \delta \, U\, U^{-1}\, \partial_j\, U\, U^{-1}\, \partial_k \, U) \nonumber\\
&=& \frac{1}{8\pi^2} \int d S^i \varepsilon^{ijk} t r (U^{-1} \, \delta \, U\, U^{-1}\, \partial_j\, U\, U^{-1}\, \partial_k \, U) = 0
\label{eqtwo7}
\end{eqnarray}
The last integral is over the surface (at infinity) bounding the base space and vanishes for localized variations $\delta U$. In fact, the entire $\omega\, (U)$, not only its variation, can be presented as a surface integral, but this requires parameterizing the group element $U\, \text{on}\, R^3$. For example for $SU(2)$,
\begin{alignat}{1}
&U = exp \lambda,\  \lambda = \lambda^a\, \sigma^a/2i\  (\vec\sigma \equiv \text{Pauli matrices})\nonumber\\
&\omega\, (U) = \frac{1}{16\pi^2}\, \int d S^i\, \varepsilon^{ijk}\, \varepsilon_{abc}\, \hat{\lambda}^a\, \partial_j\, \hat{\lambda}^b\, \partial_k\, \hat{\lambda}^c (\sin |\lambda | - |\lambda |) \nonumber\\ 
&|\lambda | \equiv \sqrt{\lambda^a \lambda^a}, \ \hat{\lambda}^a \equiv \lambda^a/|\lambda|.
\label{eqtwo8}
\end{alignat}
Specifically with $|\lambda|_{\ \overrightarrow{r\to \infty}} \ 2\pi\, n\, (\text{so that} \, U_{\ \overrightarrow{r\to \infty}}\ \pm I), \omega\, (U) = -n$.  As befits a topological entity, $\omega\, (U)$ is determined by global (here large distance) properties of $U$. 

Since all gauge transformations, small and large, are symmetry operations for the theory, (\ref{eqtwo5}) should be generalized to 
\begin{equation}
\Psi (\bfA^{U_n}) = e^{in\theta}\, \Psi (\bfA),
\label{eqtwo9}
\end{equation}
where $\theta$ is an universal constant. Thus Yang-Mills quantum states behave as Bloch waves in a periodic lattice, with large gauge transformations playing the role of lattice translations and the Yang-Mills vacuum angle $\theta$ playing the role of the Bloch momentum. This is further understood by noting that the profile of the potential energy density, $\frac{1}{2}\, \bfB^a\cdot\bfB^a$ possesses a periodic structure symbolically depicted on Figure 2.\\
\begin{figure}[t]
   \centering
   \includegraphics[scale=.90]{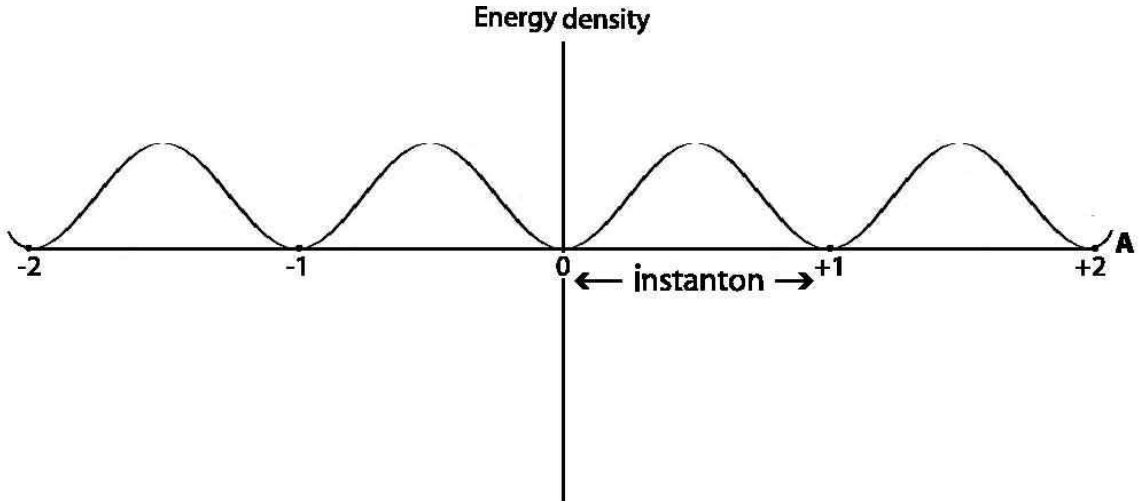} 
   \caption{Schematic for energy periodicity of Yang-Mills fields.}
   \label{ymfields}
\end{figure}
Thanks to Gauss' law, potentials $\bfA$ that differ by small gauge transformations are identified, while those differing by large gauge transformations give rise to the periodicity. Zero energy troughs correspond to pure gauge vector potentials in different homotopy classes $n$: $\bfA = - U^{-1}_n\, \vec\nabla\, U_n$.

The $\theta$ angle (Bloch momentum) arises from quantum tunneling in $\bfA$ space. Usually in field theory tunneling is suppressed by infinite energy barriers. (This gives rise to spontaneous symmetry breaking.) However, in Yang-Mills theory there are paths in field space that avoid such barriers. Quantum tunneling paths are exhibited in a semiclassical approximation by identifying classical motion in imaginary time (Euclidean space) that interpolates between classically degenerate vacua and possesses finite action.

In Yang-Mills theory, continuation to imaginary time, $x^0\to i \, x^4$, places a factor of $ i \, \text{on}\, \bfE^a$. Zero (Euclidean) energy is maintained when $\bfE^a = \pm \bfB^a$, or with covariant notation in Euclidean space
\begin{equation}
\frac{1}{2}\, \varepsilon^{\mu \nu \alpha \beta}\, F_{\alpha \beta} \equiv  \fuast = \pm\ F^{\mu \nu}.
\label{eqtwo10}
\end{equation}
 Euclidean finite action field configurations that satisfy (\ref{eqtwo10}) are called self-dual or anti self-dual instantons. By virtue of the Bianchi identity (\ref{eqone10}), instantons also solve the field equation (\ref{eqone14a}) in  Euclidean space.
Since the Euclidean action may also be written as 
\begin{equation}
I_{\scriptscriptstyle YM} = \frac{1}{4}\, \int d^4 x\, t r \, (F^{\mu \nu} \pm \fuast) (F_{\mu \nu} \pm \fdast)\, \mp \frac{1}{2} \, \int d^4 x \, t r \fuast\, F_{\mu \nu}, 
\label{eqtwo11}
\end{equation}
and the first term vanishes for instantons, we se that instantons are characterized by the last term, the Chern-Pontryagin index.
\begin{eqnarray}
\mathcal{P} &\equiv&  - \frac{1}{16\pi^2} \, \int d^4 x \, t r \, (\fuast \, F_{\mu \nu}) \nonumber\\
&=& -\frac{1}{32\pi^2} \, \int d^4 x \, \varepsilon^{\mu \nu \alpha \beta} \, t r \, ( F_{\alpha \beta}\, F_{\mu \nu})
\label{eqtwo12}
\end{eqnarray}
This again is an important topological entity: (i) The diffeomorphism invariant $\mathcal{P}$ does not involve the metric tensor. (ii) $\mathcal{P}$ is insensitive to local variations of $A_\mu$.
\begin{eqnarray}
\delta \mathcal{P} &=& - \frac{1}{8\pi^2} \, \int d^4 x \, t r (\fuast \delta\, F_{\mu \nu}) = - \frac{1}{4\pi^2} \, \int d^4x t r (\fuast\, D_\mu \delta\, A_\nu)\nonumber\\
&=& \frac{1}{4\pi^2} \, \int d^4 x\, t r \, ( D_\mu \, \fuast\, \delta \, A_v) = 0 
\label{add-in}
\end{eqnarray}
 (iii) $\mathcal{P}$ may be presented as a surface integral owing to the formula  
\begin{alignat}{1}
&\frac{1}{4}\, t r \fuast F_{\mu \nu} = \partial_\mu\, K^\mu \label{eqtwo13} \\
&K^\mu \equiv \varepsilon^{\mu \nu \alpha \beta}\, t r\, (\frac{1}{2} \, A_\alpha\, \partial_\beta\, A_\gamma + \frac{1}{3}\, A_\alpha\, A_\beta\, A_\gamma),
\label{eqtwo14}
\end{alignat}
$K^\mu$ is where the Chern-Simons current.
\begin{equation}
\mathcal{P} = - \frac{1}{4\pi^2}\, \int d \, S_\mu \, K^\mu
\label{eqtwo15}
\end{equation}

The integral (\ref{eqtwo15}) is over the base space boundary, $S^3$. The Chern-Pontryagin index of any gauge field configuration with finite (Euclidean) action (not only instantons) is quantized. This is because finite action requires $F_{\mu \nu}$ to vanish at large distances, equivalently $A_\mu \to U^{-1}\, \partial_\mu\, U$. Using this in (\ref{eqtwo14}) renders (\ref{eqtwo15}) as
\begin{equation}
\mathcal{P} = \frac{1}{24\pi^2} \, \int d S_\mu\, \varep\, t r\, (U^{-1}\, \partial_\alpha\, U\, U^{-1}\, \partial_\beta\, U \, U^{-1}\, \partial_\gamma\, U),
\label{eqtwo16}
\end{equation}
which is the  same as (\ref{eqtwo5}), and for the same reason  is given by an integer $[\Pi^3\, (G) = Z]$. Alternatively, for  instantons in the (Euclidean) Weyl gauge $(A_4 = 0)$, which interpolate as 
$x^4$ passes from $-\propto \text{to}\, + \infty$ between degenerate, classical vacua $A_i = 0\, \text{and}\, A_i = - U^{-1}\, \nabla_i\, U, \\ 
\mathcal{P}$ becomes
\begin{eqnarray}
\mathcal{P} &=& \frac{1}{4\pi^2}\, \int d x^4\, d^3 x\, (\partial_4 K^4 + \vec\nabla \cdot {\bf K}) \nonumber \\
                      &=& \frac{1}{4\pi^2} \int d^3 x \, K^4 \mid_{x^4 = \infty} \nonumber \\
                      &=& \frac{1}{24\pi^2} \int d^3 x\, \varepsilon^{ijk}\, t r\,  \U \partial_i \, \UU \partial_j \, \UU \partial U = \omega (U).
                      \label{eqtwo17}
\end{eqnarray}
We have assumed that the potentials decrease at large arguments sufficiently rapidly so that the gradient term in the first integrand does not contribute. This rederivation of  (\ref{eqtwo16}) relies on the ``motion" of an instanton between vacuum configuration of different winding number. 

An explicit 1-instanton $SU(2)$ solution $(\mathcal{P}=1)$ is
\begin{equation}
A_\mu = \frac{-2 i}{(x-\xi)^2 + \rho^2}\ \ \sigma_{\mu \nu}\, x^\nu
\label{eqtwo18}
\end{equation}
(Upon reinserting the coupling constant $g$, which has been scaled to unity, the field profiles acquire the factor $g^{-1}$.)
In (\ref{eqtwo18}), $\sigma_{\mu \nu} \equiv \frac{1}{4 i} \, (\sigma^\dagger_\mu\, \sigma_\nu - \sigma^\dagger_\mu\, \sigma_\mu), \sigma_\mu \equiv (-i\vec\sigma, I)$.
$\xi$ is the ``location" of the instanton, $\rho$ is its ``size", and there are 3 more implicit parameters fixings the gauge, for a total of 8 parameters that are needed to specify a single $SU(2)$ instanton.  One can show that there exist $N$ instanton/anti-instanton solutions $(\mathcal{P} = N/\text{-}N)$ and in $SU(2)$ they depend on $8 N$ parameters.
From (\ref{eqtwo11}) we see that at fixed $N$, instantons minimize the (Euclidean) action. Explicit formulas exist for the most general $N=2$ solution, while for $N\geq 3$ explicit formulas exhibit only $5N+7$ parameters. But algorithms have been found that construct the most general $8N$- parameter instantons. 
The 1-instanton solution is unchanged by $SO(5)$ rotations, the maximal compact subgroup of the $SO(5.1)$ conformal invariance group for the Euclidean-4 space Yang-Mills equation (\ref{eqone14a}).

The Chern-Pontryagin index also appears in the Yang-Mills quantum action, for the following reason. Since all physical states respond to gauge transformations $U_n$ with the universal phase $n \theta$ (\ref{eqtwo9}), physical states may be presented in factorized form,
\begin{equation}
\Psi (\bfA) = e^{i \mu W (\bfA) \theta}\,  \Phi (\bfA),
\label{eqtwo19}
\end{equation}
where $\Phi (\bfA)$ is invariant against all gauge transformations, small and large, while the phase response is carried by $W(\bfA)$.
\begin{equation}
W(\bfA^{U_n}) = W (\bfA) + n
\label{eqtwo20}
\end{equation}
An explicit expression for  $W (\bfA)$ is given by $-\frac{1}{4\pi^2}\, \int d^3 x\, K^0$ where $K^0$ is the time (fourth) component of $K^\mu$, with dependence on the fourth variable suppressed, {\it i.e.} $K^0$ is defined on 3-space.
\begin{equation}
W(A) = -\frac{1}{4\pi^2} \, \int d^3 x\, \varepsilon^{ijk} \, t r\, (\frac{1}{2} \, A_i\, \partial_j \, A_k + \frac{1}{3} \, A_i\, A_j\, A_k).
\label{eqtwo22}
\end{equation}
The gauge transformation properties of $W(A)$ are
\begin{eqnarray}
W(\bfA^U) = W(\bfA) &+& \frac{1}{8\pi^2} \, \int d^3 x \, \varepsilon^{ijk}\, \partial_i\, t r \, ( \partial_j \, \UU\, A_k)\nonumber \\
&+& \frac{1}{24\pi^2} \, \int d^3 x\, \varepsilon^{ijk}\, t r \, (\U\, \partial_i\, \UU \, \partial_j \UU\, \partial_k\, U).
\label{eqtwo23}
\end{eqnarray}
The middle surface term does not contribute for well behaved $\bfA$; the last term is again $\omega(U)$, the winding number of the gauge transformation $U$. Thus (\ref{eqtwo20}) is verified.

The universal gauge varying phase $e^{i\theta W (\bfA)}$, which multiplies all gauge invariant functional states, may be removed at the expense of subtracting from the action $ \theta\, \int d^4 x \ \partial_t W (\bfA) = - \frac{\theta}{4\pi^2}\, \int d^4 x\, \partial_t\, K^0 = \theta\, \mathcal{P}$ [as in (\ref{eqtwo17})]. Thus the Yang-Mills quantum action extends (\ref{eqone12}) to 
\begin{equation}
\iym^{\text{quantum}} = \int d^4 x \, t r \, (\frac{1}{2}\, \fumn \fmn + \frac{\theta}{16\pi^2}\, \fuast \fmn).
\label{eqtwo24}
\end{equation}

The additional Chern-Pontryagin term in (\ref{eqtwo24}) does not contribute to equations of motion, but it is needed to render all physical states invariant against all gauge transformations, large and small. With this transformation one sees that  the $\theta$-angle is a Lorentz invariant, but CP non-invariant effect. Evidently specifying a classical gauge theory requires fixing a group; a quantized gauge theory is specified by a group and a  $\theta$-angle, which arises from topological properties of the gauge theory. The energy eigenvalues depend on $\theta$, and distinct $\theta$ correspond to distinct theories.

Note that the reasoning leading to (\ref{eqtwo9}) and (\ref{eqtwo24}) relies on exact quantum mechanical arguments, while the instanton based tunneling discussion is semi-classical.

\subsection*{B. Adding Fermions}
When fermions couple to the gauge fields, the previously described topological effects are modified by action of the chiral anomaly. Dirac fields, either non-interacting but quantized, or unquantized but interacting with a gauge potential through a covariantly conserved current $J^\mu_a, \, \mathcal{L}_I = -J^\mu_a\, A^a_\mu$, also possess a chiral current $j^\mu_5 = \bar{\psi} \, \gamma^\mu \, \gamma_5\, \psi$, which satisfies
\begin{equation}
\partial_\mu \, j^\mu_5 = 2m\, i \bar{\psi}\, \gamma _5 \, \psi.
\label{eqtwo25}
\end{equation}
Here $m$ is the mass, if any, of the fermions. $j^\mu_5$ is conserved for massless fermions, which therefore enjoy a chiral symmetry: $\psi \to e^{i \alpha \gamma_5}\,  \psi$. However, when the interacting fermions are quantized, there arises  correction to (\ref{eqtwo25}); this is the chiral anomaly. 
\begin{equation}
\partial_\mu \jmu_A = 2 im \langle \bar{\psi} \gamma_5 \psi\rangle_A + C\  \fuasta \fmna
\label{eqtwo26}
\end{equation}
$C$ is determined by the fermion quantum numbers and coupling strengths. [For a single charged (e) fermion and an $U(1)$ gauge potential, $C = e^2/8\pi^2$.]  $\langle | \rangle_A$ signifies the fermionic vacuum matrix element in the presence of $A_\mu$. The modified equation (\ref{eqtwo26}) indicates that even in the massless limit chiral symmetry remains broken due to the anomaly, which arises with quantized fermions. 

$\jmu_A$ may also be presented as
\begin{equation}
\jmu_A = t r \, \gamma_5\, \gamma^\mu \langle \psi \bar{\psi}\rangle_A.
\label{eqtwo27}
\end{equation}
In Euclidean space $\langle \psi \bar{\psi}\rangle_A$ is the coincident-point limit of the resolvent $\Rxy$ for the Dirac equation.
\begin{equation}
\Rxy = \sum_\in \, \frac{\psi_\epsilon (x)\, \psi^\dagger_\epsilon (y)}{\epsilon + i \mu}
\label{eqtwo28}
\end{equation}
Here $\psi_\epsilon$ is an eigenfunction of the massless, Euclidean Dirac operator in the presence of the gauge field $A_\mu$.
\begin{equation}
i \, \gamma^\mu\, (\partial_\mu + A_\mu)\, \psi_\epsilon = \epsilon\, \psi_\epsilon
\label{eqtwo29}
\end{equation}
The coincident-point limit is singular, so $R$ must be regulated: $R\to R - R_{\text{Reg}}$ (we do not specify the regularization procedure).
It then follows that
\begin{eqnarray}
\partial_\mu \jmu &=& 2 i \mu \sum_\epsilon\, \frac{\psi^\dagger_\epsilon (x)\, \gamma_5\, \psi_\epsilon  \, (x)}{\epsilon + i \mu}
- t r \, \gamma_5 \, \gamma^\mu \, \partial_\mu\, R_{\text{Reg}}\nonumber\\
&=& 2 i \mu\, \sum_\epsilon\, \frac{\psi^\dagger_\epsilon (x)\, \gamma_5\, \psi_\epsilon  \, (x)}{\epsilon + i \mu} + C\ \fuasta \, F^a_{\mu \nu} .
\label{eqtwo30}
\end{eqnarray}
The first term on the right is the (Euclidean space) analog of the mass term in (\ref{eqtwo25}) or (\ref{eqtwo26}), while the second survives even after the regulators are removed, giving the anomaly $t r \, \fuast \fmn$. 

The anomaly formula (\ref{eqtwo26}), or more explicitly (\ref{eqtwo30}), is also the local form of the Atiyah-Singer index theorem, which follows after (\ref{eqtwo30}) is integral over all space: The left side integrates to zero.
The integral   of the first term on the right, $\int d x \, \psi^\ast_\epsilon\, \gamma_5 \, \psi_\epsilon$, vanishes for  $\epsilon \ne 0$ by orthogonality,  because $\gamma_5\, \psi_\epsilon$ is an eigenfunction of  (\ref{eqtwo29}) with eigenvalue $-\epsilon$. Only zero modes contribute to the $\epsilon$ sum since these can be chosen to be eigenfunctions of $\gamma_5 \, $, $n_\pm$ of them satisfying $\psi_0 = \pm \, \gamma_5 \, \psi_0$.
For a single multiplet, the normalizations work out so that
\begin{equation}
n_+ - n_- = \frac{1}{16\pi^2}\, \int d^4 x \, t r \, \fuast\, \fmn .
\label{eqtwo31}
\end{equation}
The result that the (signed) number of zero modes is the Chern-Pontryagin index is an instance of the Atiyah-Singer theorem. (In specific applications one can frequently show that $n_+\, \text{or}\, n_- $ vanishes.)
It therefore follows that in the background field of instantons, the Euclidean Dirac equation possesses zero modes.

Another viewpoint on the chiral anomaly arises within the functional integral formulation, where the exponentiated action is constructed from unquantized fields, over which the functional integration is performed. Here the classical action retains chiral symmetry $\psi\to e^{i \alpha \gamma_5}\,  \psi$, but the Grassmann fermion measure $d \psi\, d \bar{\psi}$, once it is properly regularized, looses chiral invariance and acquires the anomaly.
\begin{equation}
d\psi\, d\bar{\psi} \to d\psi\, d\bar{\psi}\, exp\, i \, C  \!\!  \int d^4 x\, \alpha\, t r\, \fuast\, \fmn
\label{eqtwo32}
\end{equation}

Evidently the chiral anomaly involves the gauge theoretic topological entity, the Chern-Pontryagin density. Not unexpectantly the anomaly phenomenon affects significantly the topological properties of the gauge theory that are connected to $\mathcal{P}$ and were described previously. 

When there is (at least) one massless fermion coupling to the Yang-Mills fields, the Yang-Mills $\theta$-angle looses physical relevance. This is because a chiral transformation that redefines the massless Dirac field does not modify the classical action, but owing to the chiral non-invariance of the functional measure, (\ref{eqtwo32}), an anomaly term is induced in the (effective) quantum action. The strength of this induced term can be fixed so that it cancels the $\theta$-term in (\ref{eqtwo24}). 
Since field redefinition cannot affect physics, the elimination of the $\theta$-term indicates that it had no physical relevance in the first place. In particular energy eigenvalues no longer depend on $\theta$.

An alternate argument for the same conclusion is based on the functional determinant that arises when the functional integral is  performed over the massless Dirac field: $\mathtt{det} [\gamma^\mu\, (\partial_\mu + A_\mu)]$. The semi-classical tunneling analysis of the $\theta$-angle is based on instantons, but in the presence of instantons the Dirac equation has a zero mode (\ref{eqtwo31}). Consequently the determinant vanishes, tunneling is suppressed and so is the $\theta$-angle.

But in the standard model for particle physics there are no massless fermions, so the  presence of the $\theta$-angle entails the following physical consequences. The tunneling amplitude $\Gamma$ in leading semi-classical approximation is determined by the Euclidean action, {\it viz.} the continuation of $i \iym$ in (\ref{eqtwo24}) to imaginary time. This results in the same expression except that the topological $\theta$-term acquires a factor of  $i$. Only the $1$-instanton and anti-instanton give the dominant contribution,
\begin{equation}
\Gamma \propto \ \cos \, \theta \, e^{- 8\pi^2/g^2},
\label{eqtwo33}
\end{equation}
where the coupling constraint $g$ has been reinserted; the proportionality constant has not been computed, owing to infrared divergences.
(Higher instanton number configurations contribute at an exponentially subdominant order and have thus far played no role in physics.) The tunneling leads to baryon decay, but fortunately at an exponentially small rate. More useful is the fact that instanton tunneling gives semi-classical evidence for the removal of an unwanted chiral $U(1)$ Goldstone symmetry, which would be present in the standard model if the chiral anomaly did not interfere. Furthermore, the chiral anomaly facilitates the decay of the neutral pion to two photons; a process forbidden by other apparent chiral symmetries of the standard model, which in fact are modified by the chiral anomaly. Gauge fields in four dimensions must interact with anomaly-free currents. This necessitates a precise adjustment of fermion content and charges so that the anomaly coefficents  [analogs of ``C" in (\ref{eqtwo26})] vanish for currents coupled to gauge fields .
Finally, $\theta$ provides a tantalizing source of CP violation in the strong interaction sector of the standard model. But no experimental signal ({\it e.g.} neutron electric dipole moment) for this effect has been seen. At present we do not know what mechanism is responsible for keeping $\theta$ vanishingly small.

These are the physical consequences of topological effects in 4-dimensional gauge theories. Although they have provided experimentalist with only a few numbers to measure ({\it e.g.} $\pi^0 \to 2 \gamma$ decay amplitude, prediction of anomaly-free arrangements of quarks and leptons in families) they have added enormously to our appreciation of the complexities of quantized gauge theories.

That chiral anomalies are an obstruction to consistent gauge interactions can be established within perturbation theory. A similar, but non perturbative effect is seen in a $SU(2)$ gauge theory with $N$ Weyl fermion $(\gamma_5\, \psi = \pm \, \psi) \ SU(2)$ doublets, which lead upon functional integration to  $\mathtt{det} \, [\gamma^\mu (\partial_\mu + A_\mu)]^{N/2}$. But because $\Pi^4 \big(SU(2)\big) = Z_2$, there exists a single homotopy class of gauge transformations which are not deformable to the identity.  One shows that the determinant changes sign when such a gauge transformation is performed. Thus the theory is ill-defined for odd $N$. Consistent $SU(2)$ gauge theories must possess an even number of Weyl fermion doublets, but such models have not found a place on physical theory.

\subsection*{C. Adding Bosons}
Instantons are finite-action solutions to classical equations continued to imaginary time; they provide a semi-classical description  of quantum mechanical tunneling. A field theory may also possess finite-energy, time-independent (static) solutions to the real-time equations of motion. When thes solutions are stable for topological reasons they are called ``solitons". Solitons give semi-classical evidence for the existence in the quantum field theory of a particle sector disjoint from the particles obtained by quantizing field fluctuations around the vacuum state. The soliton particles are heavy for weak coupling $g$. [Their energy is $0 (1/g^2)$; the field profiles are $0 (1/g)$.] They do not decay owing to the conservation of ``charges" that do not arise from Noether's theorem but are topological.

Yang-Mills theory does not possess soliton solutions (except in 5-dimensional space-time, where the static solitons are just the 4-dimensional instantons discussed previously). However, when a gauge theory, based on a simple group is coupled to  a scalar field that undergoes symmetry breaking to $U(1)$, soliton solutions exist. These are the `tHooft-Polyakov magnetic monopoles, found in a $SU(2)$ gauge theory with scalar fields in the adjoint representation, as well as various generalizations. The topological consideration that arises here concerns finite energy of the static, scalar field multiplet $\varphi$, which in the Weyl gauge is 
\begin{equation}
E (\varphi) = \int d^3 x \, \bigg( (\bfD \varphi)^a \cdot (\bfD \varphi)^a\,  |^2 + V (\varphi)\bigg)
\label{eqtwo34}
\end{equation}
$V$ is non-negative and possesses no trivial symmetry breaking zeroes. On the sphere $S^2$ at spatial infinity $\varphi$  must tend to such a zero.  Thus the fields belong to G/H, where G is the gauge group and H the unbroken subgroup. For the `tHooft-Polyakov monopole these are  $SU(2)$ and $U(1)$ respectively, and the scalar field provides a mapping of the sphere at infinity $S^2$ to $S^2 \approx SU(2)/U(1)$.

One now considers $\Pi^2\, (S^2) = \Pi^2\, (SU(2)/U(1)) = \Pi^1\, (U(1)) = Z$, and one shows that the magnetic flux is determined by the winding number. Hence the magnetic charge is quantized. Explicitly, the electromagnetic $U(1)$ gauge field is given by 
\begin{eqnarray}
f_{\mu \nu} &\equiv& \hat{\varphi}^a \, \fdasta - \varepsilon_{abc}\, \hat{\varphi}^a \, (D_\mu \hatp)^b\, (D_\nu \hatp)^c \nonumber\\
&=& \partial_\mu \, a_\nu - \partial_\nu\, a_\mu \nonumber \\
&& a_\mu \equiv \hatp^a\, A^a_\mu - \cos\, \alpha \, \partial_\mu\, \beta,
\label{eqtwo35}
\end{eqnarray}
where $\hatp^a$ is the unit iso-vector, parameterized as $\hatp^a = (\sin\alpha  \cos \beta, \sin \alpha\, \sin \beta, \cos \alpha)$.
The manifestly conserved magnetic current
\begin{subequations}
\begin{equation}
j^\mu_m = \partial_\nu \, {^\ast f}^{\mu \nu}
\label{eqtwo36a}
\end{equation}
is rearranged to read
\begin{equation}
j^\mu_m = - \frac{1}{2}\, \varep \, \varepsilon_{abc}\, \partial_\alpha\, \hatp^a\, \partial_\beta\, \hatp^b\, \partial_\gamma\, \hatp^c ,
\label{eqtwo36b}
\end{equation}
\end{subequations}
and is nonvanishing because $\hatp^a$ possesses zeroes, where $\partial_\alpha\, \hatp^a$ acquires  localized singularities. The magnetic charge
\begin{equation*}
m = -\frac{1}{4\pi}\, \int d^3 x \, j^0_m = \frac{1}{4\pi}\, \int d^3 x\, \vec\nabla \cdot \bfb
\end{equation*}
[$b^i = U(1) \, \text{magnetic field:} \, -\frac{1}{2}\, \varepsilon^{ijk}\, f_{jk} = {^\ast f}^{i0}$] is given by the topological entity (Kronecker index of the mapping)
\begin{eqnarray*}
m &=& \frac{1}{8\pi}\, \int d^3 x \, \varepsilon^{ijk}\, \varepsilon_{abc}\, \partial_i\, \hatp^a\, \partial_j \, \hatp^b \, \partial_k\, \hatp^c \\
&=& \frac{1}{8\pi}\, \int dS^i \varepsilon^{ijk}\, \varepsilon_{abc}\, \hatp^a\, \partial_j \, \hatp^b \, \partial_k\, \hatp^c \\
&=& - \frac{1}{4 \pi}\,  \int dS^i \varepsilon^{ijk}\, \partial_j \cos \alpha\, \partial_k \, \beta
\end{eqnarray*}
which readily evaluates the integer winding number.

The theory also supports charged magnetic monopole solutions called ``dyons". Here the profiles involve time-periodic gauge potentials, where the time variation is just a gauge transformation $\partial_t\, A_\mu = D_\mu \,\lambda$.  (Gauge equivalent, static expressions have slow large-distance fall-off, which is removed by the time dependent gauge function.) For dyons, the integer valued Chern-Pontryagin index, with the integration taken over all space and in time over the dyon period, reproduces the magnetic monopole strength. 

Regrettably, these fascinating structures are not found in Nature. Nor do they arise in the standard model, whose structure group is not simple, although speculative grand unified models, with simple $G$ and $H = SU(3) \times U(1)$, would support magnetic monopoles and dyons. While challenged physically, the magnetic monopole phenomena has produced extensive and interesting mathematical analysis.

\section{Gauge Theories in Two Dimensions}
Two dimensional gauge theories have few physical applications; edge states of  the planar quantum Hall effect can be described by excitations moving on a line. However, the Abelian model with fermions is useful in that it provides a very accurate reflection of topological behavior in the physically important 4-dimensional theory.
 \subsection*{A. Abelian Gauge Theory}
Take the spatial interval to be $[-L, L]$. Homotopically nontrivial gauge transformations satisfy $\lambda (L) - \lambda(-L) = 2\pi n\  [\Pi^1 (U^1) = Z].$ States $\Psi (A)$ of the free gauge theory that satisfy Gauss' law and respond with a $\theta$-angle are 
\begin{eqnarray}
\Psi (A) = \mathrm{exp}\, \frac{i\theta}{2\pi}\, \int d x\, A, \nonumber\\
\Psi (A + \partial \lambda) = e^{in\theta}\, \Psi(A).
\label{eqthree1}
\end{eqnarray}
In this model $\theta$ has the interpretation of a constant background electric field $\mathcal{E} = - \theta /2 \pi$.
\begin{eqnarray}
 E \Psi(A) \!\!&=&\!\! \mathcal{E} \Psi(A), \ E \equiv F_{01}\  \nonumber\\ 
 i\, \frac{\delta}{\delta A}\, \Psi(A) \!\!&=&\!\! - \frac{\theta}{2\pi}\, \Psi(A) 
\label{eqthree2}
\end{eqnarray}
This also gives the energy eigenvalue.
\begin{equation}
\frac{1}{2}\, \int d x  E^2\, \Psi(A) = \frac{1}{2}\, \int d x\, \mathcal{E}^2\, \Psi (A)
\label{eqthree3}
\end{equation}
The phase may be removed by adding to the Lagrangian $ -\frac{\theta}{2 \pi} \, \int d x\, \partial_t \, A$; equivalently the action becomes
\begin{subequations}\label{eqthree4}
\begin{equation}
I^{\mathrm{quantum}}_{EM} = \int d^2 x \, (-\frac{1}{4}\, \fumn \, \fmn + \frac{\theta}{4\pi}\, \varepsilon^{\mu \nu}\, \fmn ),
\label{eqthree4a}
\end{equation}
which apart from a constant is also given by a formula with the background field.
\begin{equation}
I^{\mathrm{quantum}}_{EM} = \frac{1}{2} \int d x \, (E + \mathcal{E})^2
\label{eqthree4b}
\end{equation}
\end{subequations}
Because of gauge invariance, there is only one state, annihilated by $E$ and carrying energy $\frac{1}{2} \int d x\,  \mathcal{E}^2$.
Distinct $\theta$ (different $\mathcal{E}$) correspond to distinct theories.

We recognize in (\ref{eqthree4a}) the 2-dimensional Chern-Pontryagin density, contributing a total derivative to the action,
\begin{equation}
\mathcal{P} = \frac{1}{4\pi} \int d^2 x \ \varepsilon^{\mu \nu}\, \fmn ;
\label{eqthree5}
\end{equation}
the Chern-Simons current, whose divergence is $\mathcal{P}$,
\begin{equation}
K^\mu = \frac{1}{2\pi} \, \varepsilon^{\mu \nu}\, A_\nu ;
\label{eqthree6}
\end{equation}
and the Chern-Simons term, which carries the phase of $\Psi$
\begin{equation}
\int d x \, K^0 = \frac{1}{2\pi} \, \int d x\, A.
\label{eqthree7}
\end{equation}
For Euclidean-space gauge potentials, which are given at large distance by the pure gauge $2\pi n \tan^{-1} y/x,  \mathcal{P} = n$. All this is just as in the 4-dimensional theory, except there are no instantons and no tunneling.

\subsection*{B. Adding Fermions}
The addition of massless fermions to the $U(1)$ gauge theory results in the Schwinger model of massless quantum electrodynamics in 2-dimensional space time. The equation of motion becomes
\begin{equation}
\partial_\mu \fumn = J^\nu
\label{eqthree8}
\end{equation}
with the vector current constructed from the Dirac fields as $J^\mu = \bar{\psi} \, \gamma^\mu\, \psi$. This current remains conserved in the  quantized version because it couples to the gauge field. But the axial vector current $j^\mu_5 = \bar{\psi}\, \gamma^\mu \, \gamma_5\, \psi$ acquires an anomaly that involves the Chern-Pontryagin density in (\ref{eqthree5}).
\begin{equation}
\partial_\mu\, j^\mu_5 = \frac{1}{2\pi}\, \varepsilon^{\mu \nu} \, \fmn
\label{eqthree9}
\end{equation}

The model is readily solved, and shows no $\theta$-angle (background field) dependence in physical quantities. The solution is directly obtained by combining (\ref{eqthree8}) with (\ref{eqthree9}) into a second order differential equation and using the matrix identity of 2-dimensional Dirac (=Pauli) matrices: $\varepsilon^{\mu \nu}\, \gamma_\nu \, \gamma_5 = \gamma^\mu$. It follows that
\begin{equation}
(\square + \frac{1}{\pi})\, E = 0.
\label{eqthree10}
\end{equation}
So the theory describes a free massive photon (mass squared = $1/\pi$ in units of $\hbar$ and the coupling constant, which have been scaled to unity), with no sign of a $\theta$-angle (background field).

But in parallel with 4-dimensional behavior, the model with massive fermions regains a $\theta$ dependence in the particles' energy spectrum; a result that is established perturbatively, because a complete solution is not available.

Note that in the Schwinger model, the gauge particle (``photon") acquires a mass, even though local gauge invariance is preserved. This happens essentially for topological/anomaly reasons. Such topological mass generation is met again in three dimensions.

\subsection*{C. Adding Bosons}
Scalar electrodynamics with a negative mass squared term in (3+1)-dimensional space-time leads to the Higgs mechanism and short range interactions due to the massive photons. In $(1+1)$ space-time dimensions, the model possesses instantons --- scalar and gauge field profiles that solve the imaginary-time equations of motion --- labeled by $\Pi^1 (U (1)) =Z$. These disorder the Higgs condensate so that the force between charged particles remains long-range, like in the positive mass squared case. This is a vivid example of how excitations arising from non-trivial topological issues significantly effect physical content.

\section{Gauge Theories in Three Dimensions}
Gauge theories  on three dimensional space-time, {\it i.e.} evolving on a plane, have physical application to planar phenomena, like the quantum Hall effect. Also the high-temperature limit of 4-dimensional field theories is governed by the corresponding field theory in three Euclidean dimensions. 

In three (more generally, odd) dimensions there are no Chern-Pontryagin quantities, no Chern-Simon currents, no axial vector currents or anomalies (there is no $\gamma_5$ matrix). These are replaced by odd-dimensional entities that can modify Yang-Mills dynamics.

\subsection*{A. Yang-Mills and other Gauge Theories}

Using the 3-index Levi-Civita tensor one can construct a gauge covariant, covariatly conserved vector, which can be added to the Yang-Mills equation. Thus (\ref{eqone14}) can be modified to 
\begin{subequations}\label{eqfour1}
\begin{equation}
D_\mu\, \fumn + \frac{m}{2}\, \varepsilon^{\mu \alpha \beta}\, F_{\alpha \beta} = J^\nu ,
\label{eqfour1a}
\end{equation}
or equivalently in terms of the dual field strength $\astf^\mu \equiv \frac{1}{2}\,  \varepsilon^{\mu \alpha \beta} \, F_{\alpha \beta}$
\begin{equation}
\varepsilon^{\mu \nu \alpha}\, D_\mu \astf_\alpha + m \, \astf^\nu = J^\nu .
\label{eqfour1b}
\end{equation}
\end{subequations}
For dimensional balance  $m$ carries dimension of mass. Indeed in the source-free case (\ref{eqfour1}) implies
\begin{equation}
(D^\alpha D_\alpha + m^2) \astf_\mu = \varepsilon_{\mu \alpha \beta}\, [\astf^\alpha, \astf^\beta].
\label{eqfour2}
\end{equation}
This shows that excitations are massive, even though local gauge invariance is preserved. Otherwise as in the Dirac monopole case, the equations of motion are unexceptional. 

However, for the quantum theory we need the action, whose variation produces the mass term in (\ref{eqfour1}). This is just the Chern-Simons term $W(A)$ in (\ref{eqtwo22}), multiplied by $-8 \pi^2 m$ and now defined on (2+1)-dimensional space-time.
\begin{equation}
 I_{CS} = 2m \, \int d^3 x \, \varepsilon^{\alpha \beta \gamma}\, t r \, (\frac{1}{2}\, A_\alpha\, \partial_\beta\, A_\gamma + \frac{1}{3}\, A_\alpha\, A_\beta\, A_\gamma)]
\label{eqfour3}
\end{equation}
Everything holds also in the Abelian theory; the last term in (\ref{eqfour3}) is then absent.

In this model the mass is generated by a topological mechanism since $I_{CS}$ possesses the usual attributes for a topological entity: It is diffeomorphisms invariant without a metric tensor; when the potentials are appropriately parameterized, it is given by a surface term. (In the Abelian case the appropriate parameterization is in terms of Clebsch decomposition, $A_\mu = \partial_\mu\, \theta + \alpha\, \partial_\mu \, \beta.$) Most importantly, in the non-Abelian theory (\ref{eqfour3}) changes by $8\, \pi^2\, m n$ with 3-dimensional gauge transformations carrying winding number $n$. Hence for consistency of the non-Abelian quantum theory $m$ must be quantized as $n/4\pi$ (in units of $\hbar$ and the coupling constant, which have been scaled to unity). All this is a clear field-theoretic analog to the quantum mechanics of the Dirac monopole, and just as for the magnetic monopole, a Hamiltonian argument for quantizing $m$ can be constructed, as an alternative to the above action-based derivation.

The time component of (\ref{eqfour1}) relates the electric and magnetic fields to the charge density.
\begin{equation}
\bfD  \cdot \bfE - m \, B = \rho 
\label{eqfour4}
\end{equation}
In the Abelian case, the first term involves a total derivative and its spatial integral vanishes, leaving a formula that identifies magnetic flux with a total charge. At low energy, the mass term dominates the conventional kinetic term in (\ref{eqfour1}), and the flux-charge relation becomes a local field-current identity.
\begin{equation}
m \, \astf^\nu \approx J^\nu 
\label{eqfour5}
\end{equation}
These formulas have made Chern-Simons-modified gauge theories relevant to issues in condensed matter  physics, for example the quantum Hall effect. In the Abelian case $m$ need not be quantized.

\subsection*{B. Adding Fermions}
3-dimensional Dirac matrices are minimally realized by 2x2 Pauli matrices. This has the consequenced that a mass term is not parity invariant; also there is no $\gamma_5$ matrix, since the product of the three Dirac ( = Pauli) matrices is proportional to $I$. While there are no chiral anomalies, there is the so-called parity anomaly: Integrating a single doublet of massless $SU(2)$ fermions one obtains $ \triangle (A)\equiv \mathsf{det} [\gamma^\mu (i \partial_\mu + A_\mu)]$, which should preserve parity and gauge invariance.

Since there are no anomalies in current divergences, $\triangle (A)$ is certainly invariant against infinitesimal gauge transformations. But for finite gauge transformations (categorized by $\Pi^3 (SU(2) = Z)$ one finds that $\triangle(A)$ is not invariant: When the gauge transformation belongs to an odd-numbered homotopy class, $\triangle(A)$ changes sign. To regain gauge invariance one must either work with an even number of fermion doublets. If only one doublet (more generally odd number) is to be used, one must add to the gauge Lagrangian a parity-violating Chern-Simons term with half the correctly quantized coefficient, to neutralize the gauge non-invariance of $\triangle(A)$.

Alternatively $\triangle(A)$ can be regularized in gauge invariant manner. But this requires massive, Pauli-Villars regulator fields, which produce a parity-violating expression for $\triangle(A)$. One cannot avoid the parity anomaly.

\subsection*{C. Adding Bosons}
There is a variety of bosonic field models that one may consider: Abelian or non-Abelian; with conventional kinetic term or supplemented by the Chern-Simons topological mass; or for low energy no kinetic term but only the Chern-Simons term, as in (\ref{eqfour5}).
Abelian charged Bose fields in a Maxwell theory lead to vortex solitons, based on $\Pi^1(U(1)) = Z$. These are just the instantons of the (1+1) dimensional bosonic gauge theory discussed previously. With Maxwell kinematics there are no charged vortices, but these appear when the Chen-Simons mass is added; see (\ref{eqfour4}). Pure Chern-Simons kinematics, with no Maxwell term can produce completely integrable soliton equations, (Liouville, Toda) when the Bose field dynamics is appropriately chosen.

\section{Conclusion}
Topological effects in field theory are associated with the infinities and regularization that beset quantum field theories. These give rise to the chiral anomaly, parity anomaly (and scale symmetry anomalies, not discussed here). Yet the anomalies themselves are finite quantities that have topological significance (Atiyah-Singer, Chern-Pontryagin, Chern-Simons). This paradoxical pairing has not been understood. Nor can we explain why the anomalies interfere in a topological manner with symmetries associated with masslessness.

Although the range of topological effects in gauge theory is large, and even larger in non-gauge theories (sigma models, Skyrme models) the relevance to actual fundamental physics is confined to the $\theta$-angle phenomenon, which is analyzed accurately and abstractly by reference to $\Pi^3(G)$ and to the interplay with fermions through the chiral anomaly. Instantons are relevant only to  an approximate, semi-classical discussion. Although after much mathematical work, general instanton configurations are well-understood, only the 1-instanton solution enjoys physical significance.

Other topological entities that fascinate are either non-existent in fundamental physics or are relevant to condensed matter physics (vortices, Chern-Simons effects). But here too, we note that the fundamental equation of condensed matter physics -- the many body Schr\"{o}dinger equation -- carries no evident topological structure. Only the phenomenological equations, which replace the fundamental one, give rise to topological intricacies.  

\newpage
\subsection*{Further Reading}
Adler, SL (1970) Perturbation Theory Anomalies.  In: Deser, S, M Grisaru, and 
H Pendleton (eds.), Lectures on Elementary Particles and Quantum 
{}Field Theory, vol. 1, pp. 3-164.  M. I. T. Press, Cambridge, MA. \\
[3ex]
Adler, SL (2004).  Anomalies to all Orders.  ArXiv: hep-th/0405040.  
To appear in: \\ 'tHooft, G (ed.) Fifty Years of Yang-Mills Theory. 
World Scientific, Singapore.\\
[3ex]
Bertlmann, RA (1996)  Anomalies in Quantum Field Theory. Clarendon Press, 
Oxford.\\
[3ex]
Coleman, S (1985) Classical Lumps and their Quantum Descendants and The Uses of Instantons. In: S. Coleman, Aspects of Symmetry, pp. 185-350. Cambridge University Press, Cambridge UK.\\
[3ex]
{}Fujikawa, K and H Suzuki (2004)  Path Integrals and Quantum Anomalies.  
Oxford University Press, Oxford.\\
[3ex]
Jackiw, R (1977) Quantum Meaning of Classical Field Theory. {\it Rev. Mod. Phys.} {\bf 49}, 681-706.\\
[3ex]
Jackiw, R (1979) Introduction to the Yang-Mills Quantum Theory. {\it Rev. Mod. Phys.} {\bf 52}, 661-673.\\
\noindent
Jackiw, R (1985)  Field Theoretic Investigations in Current Algebra and   
Topological Investigations in Quantum Gauge Theories. 
In:  Treiman, S, R Jackiw, B Zumino and E Witten,  Current Algebra 
and Anomalies, pp. 81-359.    Princeton University Press, Princeton and 
World Scientific, Singapore.\\
[3ex]
Jackiw, R and S-Y Pi (1992) Chern-Simons Solitons. {\it Prog. Theor. Phys. Suppl.} {\bf 107}, 1-40.\\
[3ex]
Jackiw, R (1995) Diverse Topics in Theoretical and Mathematical Physics. World Scientific, Singapore.\\
[3ex]
Jackiw, R. (2004) Fifty Years of Yang-Mills Theory and Our Moments of 
Triumph. ArXiv:  physics/0403109.     
To appear in:  'tHooft, G (ed.) Fifty Years of Yang-Mills Theory. 
World Scientific, Singapore.\\
[3ex]
Rajaraman, R (1982) Solitons and Instantons. North Holland, Amsterdam.\\
[3ex]
Shifman, M (1994) Instantons in Gauge Theories. World Scientific, Singapore.\\
[3ex]
`tHooft, G (1976) Symmetry Breaking through Bell-Jackiw Anomalies. {\it Phys. Rev. Lett.} {\bf 37}, 8-11.\\
[3ex]
Weinberg, S (1996)  The Quantum Theory of Fields, vol. II  chapt. 22 and 23. Cambridge University Press, Cambridge UK.

\newpage
\subsection*{Key Words}
Anomalies

axial vector

chiral 

parity

scale

\noindent
Atiyah-Singer\\
Chern-Pontryagin\\
Chern-Simons\\
CP\\
Current\\
\indent anomalous\\
\indent gauge field coupled\\
Gauge\\
\indent field\\
\indent potential\\
\indent transformation\\
Hall effect\\
Homotopy\\
Instanton\\
Kronecker index\\
 Magnetic monopole\\
\indent Dirac\\
\indent `tHooft-Polyakov\\
Mass\\
\indent topological\\
Neutron dipole moment\\
Pion decay\\
Schwinger model\\
Skyrme model\\
Solitons\\ 
$\theta$-angle\\
Tunneling\\
Vortices\\
Winding number\\
Yang-Mills






\end{document}